%% file: bare_conf_compsoc.tex
\documentclass[conference,review]{IEEEtran}

\ifCLASSOPTIONcompsoc
  \usepackage[nocompress]{cite}
\else
  \usepackage{cite}
\fi
\ifCLASSINFOpdf
\else
\fi

\usepackage{hyperref}
\usepackage{pifont}
\usepackage{spverbatim}

\usepackage{multirow}
\usepackage[normalem]{ulem}
\useunder{\uline}{\ul}{}
\usepackage{subfigure}
\usepackage[ruled,vlined,linesnumbered]{algorithm2e}
\SetKwComment{Comment}{$\triangleright$\ }{}
\usepackage{graphicx}
\usepackage{caption}
\usepackage[skip=1pt]{caption}
\usepackage{float}
\usepackage[flushleft]{threeparttable}
\usepackage{array,booktabs,makecell}
\usepackage{graphicx}
\usepackage{enumitem}
\usepackage[framemethod=default]{mdframed}
\usepackage{mathtools,lipsum}

\usepackage{cuted}
\usepackage[toc,page]{appendix}
\usepackage{balance}
\usepackage{footmisc}
\usepackage{tabularx}
\usepackage{tikz}
\usepackage{xcolor}
\usepackage{multirow}
\usepackage[many]{tcolorbox}
\usepackage{subcaption}
\usepackage[export]{adjustbox}
\usepackage{adjustbox}

\usepackage{booktabs}
\usepackage{multirow}
\usepackage{tabularx}
\usepackage{balance}
\usepackage{placeins}
\usepackage{subcaption}
\usepackage{enumitem}
\usepackage{tikz}
\usepackage{pgfplots}
\usepackage{rotating}
\usepackage[export]{adjustbox}
\usepackage[accsupp]{axessibility}

\usetikzlibrary{tikzmark,patterns,shapes.misc,shapes.geometric,positioning}
\pgfplotsset{compat=1.17}
\setlist[itemize]{itemsep=0pt, topsep=0pt}

\usepackage{listings}

\usepackage{colortbl}%

\usepackage{tikz}

\definecolor{darkred}{HTML}{860000}
\definecolor{darkteal}{HTML}{005959}
\definecolor{darkpurple}{HTML}{590059}
\definecolor{darkgrey}{HTML}{434343}

\newtcolorbox{mybox}[2][]{text width=0.95\linewidth,fontupper=\normalsize,
fonttitle=\bfseries\sffamily\scriptsize, colbacktitle=darkgrey,enhanced,
attach boxed title to top left={yshift=-2mm,xshift=3mm},
boxed title style={sharp corners},top=4pt,bottom=2pt,left=2pt,right=2pt,
  title=#2,colback=white}

\newcommand{\tool}{\textsc{Ethan}}

\begin{document}
%
\title{Image-Based Geolocation Using Large Vision-Language Models}



%


\author{\IEEEauthorblockN{Yi Liu\IEEEauthorrefmark{1}, 
Junchen Ding\IEEEauthorrefmark{2},
Gelei Deng\IEEEauthorrefmark{1}, 
Yuekang Li\IEEEauthorrefmark{2}, 
Tianwei Zhang\IEEEauthorrefmark{1}, 
Weisong Sun\IEEEauthorrefmark{1}, \\
Yaowen Zheng\IEEEauthorrefmark{3}, 
Jingquan Ge\IEEEauthorrefmark{1},
Yang Liu\IEEEauthorrefmark{1}}
\IEEEauthorblockA{\IEEEauthorrefmark{1}Nanyang Technological University, Singapore}
\IEEEauthorblockA{\IEEEauthorrefmark{2}University of New South Wales, Australia}
\IEEEauthorblockA{\IEEEauthorrefmark{3}Institute of Information Engineering, Chinese Academy of Sciences, China}}

\thispagestyle{plain}
\pagestyle{plain}
\maketitle

\begin{abstract}

Geolocation is now a vital aspect of modern life, offering numerous benefits but also presenting serious privacy concerns. The advent of large vision-language models (LVLMs) with advanced image-processing capabilities introduces new risks, as these models can inadvertently reveal sensitive geolocation information. This paper presents the first in-depth study analyzing the challenges posed by traditional deep learning and LVLM-based geolocation methods. Our findings reveal that LVLMs can accurately determine geolocations from images, even without explicit geographic training.

To address these challenges, we introduce \tool{}, an innovative framework that significantly enhances image-based geolocation accuracy. \tool{} employs a systematic chain-of-thought (CoT) approach, mimicking human geoguessing strategies by carefully analyzing visual and contextual cues such as vehicle types, architectural styles, natural landscapes, and cultural elements. Extensive testing on a dataset of 50,000 ground-truth data points shows that \tool{} outperforms both traditional models and human benchmarks in accuracy. It achieves an impressive average score of 4550.5 in the GeoGuessr game, with an 85.37\% win rate, and delivers highly precise geolocation predictions, with the closest distances as accurate as 0.3 km. Furthermore, our study highlights issues related to dataset integrity, leading to the creation of a more robust dataset and a refined framework that leverages LVLMs' cognitive capabilities to improve geolocation precision. These findings underscore \tool{}'s superior ability to interpret complex visual data, the urgent need to address emerging security vulnerabilities posed by LVLMs, and the importance of responsible AI development to ensure user privacy protection.
\end{abstract}


%
\IEEEpeerreviewmaketitle

\input{Tex/1-Introduction}
\input{Tex/2-Background}
\input{Tex/3-Study}

\input{Tex/4-Methodology}

\input{Tex/5-Evaluation}
\input{Tex/6-Discussion}

\input{Tex/7-Conclusion}

\clearpage
\bibliographystyle{IEEEtran}

\bibliography{sample-base}


\end{document}

%% file: Tex/1-Introduction.tex
\section{Introduction}\label{sec:intro}
Geolocation~\cite{luo2022geolocation, muller2018geolocation, djuknic2001geolocation, shavitt2011geolocation} is a crucial aspect of user privacy. With the growing ubiquity of smartphones and other mobile devices, image sharing has gained increasing popularity on social networks like Facebook~\cite{facebook}, Instagram~\cite{instagram}, and Foursquare~\cite{foursquare2024}. During social image sharing, privacy protection has become a critical issue because images can reveal when and where a special moment took place, who participated, and what their relationships were. Sharing images can expose much of users’ personal and social environments and their private lives. There have been multiple incidents~\cite{brumfiel2023, yang2020protecting} where people were fired due to their private photos being disclosed to unwanted audiences. The ability to accurately determine the geolocation from a photo, i.e., the ``image-based geolocation'', has significant implications for security, navigation, and social media, making geolocation a vital privacy concern.

\begin{figure}
    \centering
    \includegraphics[width=1\linewidth]{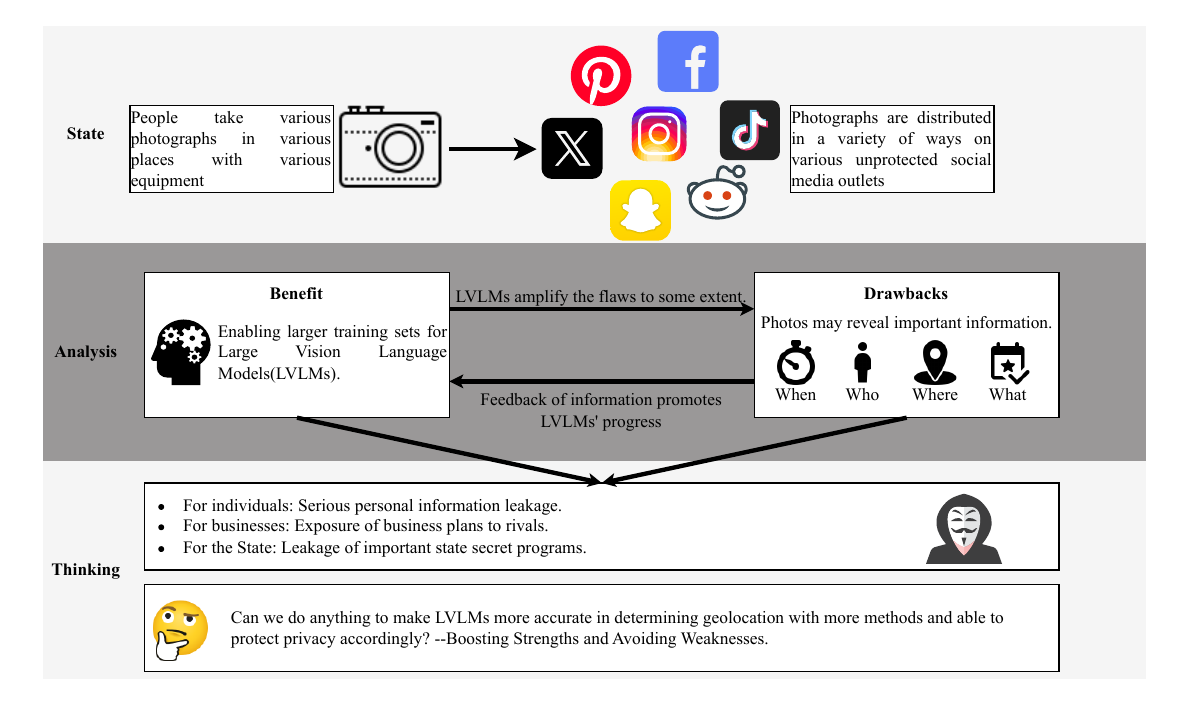}
    \caption{Common scenarios of how adversaries extract private geolocation information from the victims.}
    \label{fig:geolocation-privacy-example}
\end{figure}

With more images being associated with geo-tags and timestamps, image privacy is now closely tied to location privacy. Such exposure can have undesired consequences, especially when individuals visit sensitive locations, as illustrated in Figure~\ref{fig:geolocation-privacy-example}. For instance, consider a journalist named Alice attending a confidential meeting with a whistleblower at a coffee shop. If another patron, Tom, takes a photo of the coffee shop with Alice and the whistleblower in the background and posts it online, it could jeopardize Alice’s safety and compromise the anonymity of the whistleblower. Similar scenarios can occur in sensitive locations like political rallies or health clinics, causing significant distress if photos are shared online. Attackers could also piece together a person’s travel route by analyzing unprotected online photos, using face recognition and extracting location and timestamp information through geo-tags, metadata, or landmarks identified by advanced image processing tools. In Section 2, we demonstrate an example of such an attack to show its feasibility.

Existing AI techniques, such as those exemplified by models like GeoSpy~\cite{geospyai10:online}, have been effective in geolocation tasks, leveraging advanced capabilities to accurately determine locations from images and providing detailed reasoning behind their analyses. These models, however, often require specific technical expertise and extensive setup, limiting their accessibility to users with specialized knowledge.
In contrast, the rise of large vision language models~\cite{zhu2023minigpt, zhou2022learning, zhou2023analyzing, zhang2024vision} (LVLMs) introduces a significantly more potent capability that could potentially threaten geolocation privacy. Known for their advanced proficiency in complex tasks such as object recognition~\cite{chen2024taskclip}, scene interpretation~\cite{de2023semantic}, and content moderation~\cite{kumar2023watch}, LVLMs make it substantially easier and more accessible for anyone, even those without prior technical knowledge, to predict user locations from photos. This accessibility is largely due to LVLMs' ability to integrate and interpret extensive and diverse data types effortlessly. One example is that users create their own ChatGPT-based application for users to upload images and identify its geolocation~\cite{geolocator}.Consequently, while LVLMs expand the possibilities for user-friendly applications, they also amplify concerns about privacy and the potential for misuse.

Building on the discussion of the accessibility and potential threats posed by LVLMs in geolocation tasks, it is notable that, despite these advancements, there remains a significant gap in comprehensive evaluations of how these models perform in extracting geolocation from given photos. To date, most research works~\cite{haas2024pigeon, wazzan2024comparing, mendes2024granular} have focused on isolated aspects of geolocation or specific datasets, which fails to provide a holistic assessment of models’ overall capabilities, limitations, and practical implications. This lack of systematic study is particularly concerning given the ease with which LVLMs with image modality can be deployed by individuals without specialized knowledge, potentially amplifying privacy risks.
This gap underscores the urgent need for systematic studies that not only assess the accuracy of these models but also explore factors influencing their performance and potential security risks. Such research is vital for advancing the field responsibly and ensuring that the deployment of these powerful tools does not inadvertently compromise user privacy. Addressing these concerns is crucial for developing guidelines and technologies that safeguard against the misuse of AI in sensitive applications like geolocation.

\begin{figure}
    \centering
    \includegraphics[width=1\linewidth]{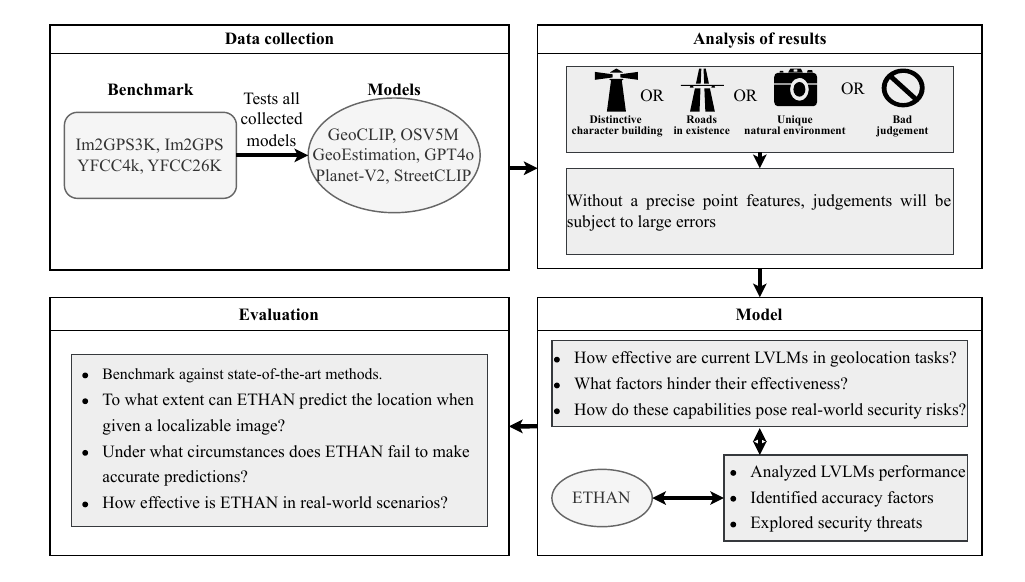}
    \caption{Overview of our work}
    \label{fig:enter-label}
\end{figure}

In this paper, we conduct the first comprehensive study that systematically evaluates the performance of existing LVLMs in extracting geolocation from photos and compares them with traditional geolocation solutions. We aim to answer three key research questions: (1) \textit{How effective are current LVLMs in geolocation tasks?} (2) \textit{Can LVLMs be applied or enhanced to exploit user privacy under real-world settings?} (3) \textit{What factors affect their proficiency in geolocation tasks?} To address these questions, we designed a comprehensive experimental framework that incorporates diverse datasets, robust evaluation metrics, and a thorough analysis of the models' outputs. This framework helps us pinpoint strengths and weaknesses in the models' performance and provides insights into potential improvements for better accuracy and reliability.

Our study provides a detailed analysis of the performance of state-of-the-art LVLMs in geolocation tasks, identifying factors that affect their accuracy and exploring potential security threats posed by their capabilities. We find that LVLMs can accurately provide geoposition information for certain images; however, their performance is heavily influenced by the quality and diversity of the data, the presence of distinctive landmarks in the photos, and the models' ability to generalize from known to unknown locations. Additionally, our findings indicate that images from urban environments, rich with identifiable features, are more likely to yield accurate geolocation results compared to those from rural or less distinctive settings. We also reveal that while LVLMs are adept at recognizing and understanding objects within images, their capabilities are largely dependent on ``knowledge'' of landmarks rather than on reasoning abilities akin to those of human geoguessing experts, who utilize a broader array of contextual cues such as landscape, weather, buildings, and vegetation to deduce locations.

Inspired by the strategies employed by human geoguessing experts, we propose \tool{}, a comprehensive framework that enhances geolocation accuracy by embedding LVLMs' reasoning capabilities into the process. We began by creating a non-biased dataset, removing inherently biased images, such as indoor scenes, from existing datasets. We then extract key objects from these images, and use them along with the image and their location ground truth to fine-tune a large vision language model. A Chain-of-Thought~\cite{wei2023chainofthought, zhang2023multimodal} (CoT) based strategy is designed, prompting the large vision language model to mimic the strategies of human experts by identifying key information from the images that may aid the geolocation process and making the guesses step-by-step. 

To evaluate the effectiveness of \tool{}, we have conducted a comprehensive evaluation. \tool{} has been tested using a dataset of approximately 50,000 ground-truth data points, split into 30,000 for training and 20,000 for testing, to ensure robust assessment. \tool{} has demonstrated superior performance across all metrics, with an average score of 4550.5 in the GeoGuessr game, significantly higher than the human average of 4120.3. It has achieved a win rate of 85.37\% and exhibited high precision, with closest distance predictions as accurate as 0.3 km and farthest distances up to 5200.2 km. These results have highlighted \tool{}'s advanced capabilities in interpreting complex visual data and accurately predicting geographical locations, outperforming other models and human competitors alike, and proving its effectiveness in both simulated and real-world scenarios.

To summarize, our key contributions in this manuscript are as follows:
\begin{enumerate}
    \item \textbf{Comprehensive Evaluation and Security Implications:} We perform the first systematic evaluation of the capabilities of existing LVLMs in extracting geolocation from photos, employing diverse datasets and robust evaluation metrics to ensure a thorough assessment. This evaluation also includes an exploration of potential security threats posed by the misuse of LVLMs in geolocation tasks, discussing real-world vulnerabilities and proposing strategies to mitigate privacy risks.
    \item \textbf{Performance Analysis:} We identify and analyze factors influencing LVLM performance in geolocation tasks, particularly highlighting the impact of data quality, diversity, the presence of distinctive landmarks, and the models’ generalization abilities across different environments.
    \item \textbf{Geolocation Framework:} We introduce \tool{}~\cite{geolocation-privacy}, a novel framework that significantly enhances the accuracy of geolocation predictions by integrating LVLMs’ reasoning capabilities to mimic human geoguessing expertise. This framework not only improves performance on benchmark datasets, often outperforming existing solutions, but also demonstrates its effectiveness in competitive environments like GeoGuessr.
\end{enumerate}

\textbf{Ethical Considerations \& Disclosure.} We acknowledge the ethical implications of our work and have taken steps to ensure that our research adheres to ethical standards. We discuss the potential privacy concerns and provide recommendations for mitigating risks associated with geolocation extraction. Our study emphasizes the importance of responsible AI development and the need for robust privacy protections to prevent misuse of geolocation technologies.

%% file: Tex/2-Background.tex
\section{Background}
\subsection{Location-dependent Privacy}
A photograph can inadvertently expose a person's location in multiple ways. Even though social media platforms like Facebook~\cite{facebook} and Instagram~\cite{instagram} remove metadata like EXIF~\cite{alvarez2004using} (Exchangeable Image File Format), which contains date and GPS coordinates that are created when photos are taken, from the uploaded photos, they store this information in a separate database. If a hacker accesses these databases, they can track users more efficiently by examining the stored metadata instead of having to extract or analyze the images individually. Besides metadata, the image itself can reveal location details through visible landmarks and street signs. Furthermore, crowdsourcing, where locals recognize and point out familiar locations shown in the photographs, also compromises privacy. Simply stripping metadata from photos is insufficient to protect the location privacy of individuals depicted in those images~\cite{brumfiel2023}.

\subsection{Localizability of Images}
The ease with which an image can be localized varies significantly based on its content and features. Images that are highly localizable often lack contextual background or are of such low quality that precise location identification becomes challenging. The more detailed and clear an image is, the more likely it is to contain explicit hints that aid in localization, such as recognizable people, animals, objects in indoor settings, distinct natural formations, specific urban structures, or historical landmarks. Images captured from street views are typically easier to localize than those from less distinctive, rural backgrounds~\cite{yang2020protecting}. In this analysis, a pivotal aspect is the interplay between the volume of available data and regulatory frameworks. At one end of the spectrum, highly localizable but easily interpretable images, like advertisements, immediately connect an object with its geographical surroundings.

\begin{figure*}[t!]
    \centering
    \includegraphics[width=1\linewidth]{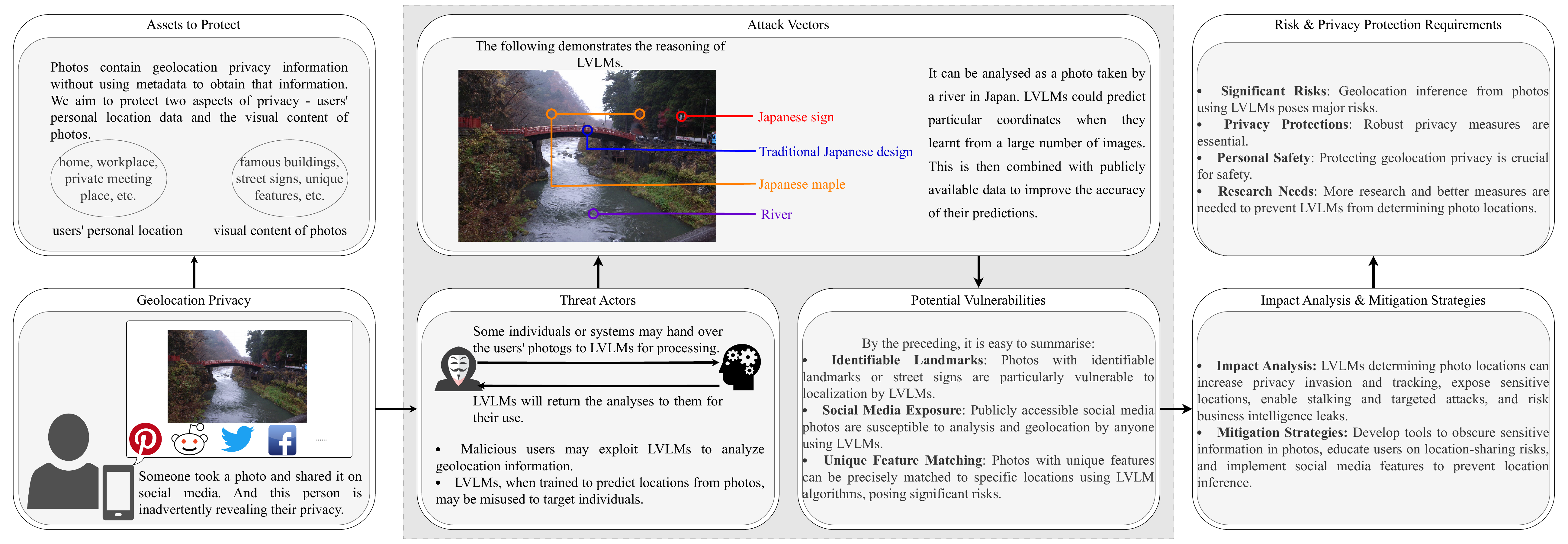}
    \caption{Threat Model for Geolocation Privacy using Large Vision Language Models (LVLMs)}
    \label{fig:threat-model}
\end{figure*}

\subsection{Overview of Geolocation Techniques}

Place recognition and visual localization are established tasks where the goal is to determine the exact position of images within a known environment. In contrast, visual geolocation is aimed at predicting 2D coordinates or identifying general locations such as countries, with a focus on achieving broad applicability and moderate accuracy, even in areas not previously seen in the training data. Methods for geolocating images generally fall into one of several categories based on whether they approach the task as an image retrieval challenge, a classification problem, or a combination of both, along with approaches utilizing LVLMs.

\noindent\textbf{Image Retrieval-Based Methods.} One intuitive approach to image geolocation is to match an image with the most similar one in a vast database, and then use the known location of that image as the prediction. Early methods~\cite{mousavian2016semantic, kalogerakis2009image, liu2014your} achieved this by matching based on simple features such as color histograms, GIST features, or textural elements. Enhancements were later made by incorporating SIFT features and support vector machines~\cite{jones2012image}. The advent of deep learning further refined these methods by employing deep features, significantly improving performance. Although these models can be highly effective with a sufficiently comprehensive and detailed image database, they do not involve learning to represent the images themselves. As a result, their effectiveness is limited in areas that are not well-represented or that change over time.

\noindent\textbf{Classification-Based Methods.} Alternatively, geolocation can be treated as a classification task by dividing the world into discrete geographic cells, based on latitude and longitude coordinates. The design of these partitions is crucial and can vary widely, including regular, adaptive, semantically driven, combinatorial, administrative, and hierarchical formats. Classification approaches must carefully balance the number and size of these cells. Overly broad cells can reduce accuracy, whereas too many small cells may not provide enough data for effective learning. Moreover, typical classification losses, such as cross-entropy, do not consider the geographical proximity of the cells; hence, mistaking two neighboring areas might be as severe as confusing two widely separated continents.

\noindent\textbf{Hybrid Approaches.} To address the limitations associated with simple discretization, some methods~\cite{chen2015camera, haas2024pigeon} combine retrieval and classification strategies. This integration is often implemented through ranking losses or contrastive objectives. For example, one strategy begins with classification and follows with regression using prototype networks. Additionally, extending the scope of predictions by estimating probability distributions over locations using spherical Gaussians is another innovative approach.

\noindent\textbf{LVLM-based Methods.} Large vision language models (LVLMs), especially based on transformer architectures, have shown significant potential in identifying geographic locations from images. It is reported that the state-of-the-art LVLM solution GPT-4~\cite{gpt4} has been used on geolocation~\cite{brumfiel2023}, and people have created GPT applications on top of the models~\cite{geolocator}. These models with image modality capability analyze visual content and contextual cues within an image to infer locations. For example, by processing elements such as text in signage, architectural styles, and natural landmarks, LVLMs can generate hypotheses about where a photo might have been taken. This capability introduces a new layer of complexity in managing location-dependent privacy, as LVLMs can bypass traditional privacy protections that focus merely on removing or obscuring metadata.

\section{Motivation and Threat Model}\label{sec:threat-model}

We illustrate our key motivation and concrete threat model in Figure~\ref{fig:threat-model}. Central to this work is the protection of geolocation privacy, which is crucial as it shields details of an individual's location and movements—details often inadvertently exposed through online photo sharing. This study focuses on how LVLMs can extract geolocation data from photos based solely on visual content, thus presenting significant privacy risks in the absence of metadata. The motivation behind this task is to underscore the urgent need for robust protections to secure geolocation privacy. The pervasive use of LVLMs to analyze and infer locations from photos highlights the critical need for advancing research and developing improved privacy measures to protect individuals from potential misuses of this technology.

\noindent\textbf{Assets to Protect:} At the heart of our concerns are two main assets: users' personal location data and the visual content of photos. Personal location data includes sensitive places such as homes, workplaces, or private gathering venues. The visual content encompasses identifiable elements such as landmarks, street signs, and unique features, all of which can potentially disclose a photo's location.

\noindent\textbf{Threat Actors:} Various actors exploit these vulnerabilities, ranging from malicious individuals to automated systems. These actors deploy LVLMs to analyze the visual content of photos and infer locations, using these insights for potentially harmful purposes.

\noindent\textbf{Attack Vectors:} The primary method employed by these threat actors involves sophisticated image analysis techniques. LVLMs, trained on diverse image datasets, excel at recognizing and interpreting visual cues such as architectural styles, signage, and natural landscapes to pinpoint locations. This capability is often enhanced by correlating LVLM inferences with publicly available geographic data, thus improving the accuracy of location predictions.

\noindent\textbf{Potential Vulnerabilities:} This capability exposes several vulnerabilities. Photos that feature clear, identifiable landmarks or unique geographical features are particularly prone to localization. The widespread availability of publicly shared photos on social media platforms amplifies this risk, providing a readily accessible source of data for analysis. Moreover, images captured in unique settings can be precisely matched to specific locations using LVLM algorithms, significantly increasing the threat landscape.

\noindent\textbf{Impact and Mitigation Strategies:} If these vulnerabilities are exploited, they could lead to severe privacy breaches. The ability of LVLMs to determine locations from photos can facilitate tracking and surveillance, exposing sensitive personal and professional locales. Such breaches not only endanger individual privacy and safety but can also lead to targeted attacks, stalking, and, in business contexts, unintended disclosures of strategic information.

Addressing these risks requires a comprehensive approach. Developing tools that leverage LVLM capabilities to detect and obscure sensitive information in photos is crucial. It is equally important to educate users about the dangers of sharing geolocation-rich photos and to promote cautious sharing practices. Furthermore, social media platforms should enhance their systems to detect and filter out images that could inadvertently reveal location information, thus preventing malicious actors from exploiting this data.

%% file: Tex/3-Study.tex
\begin{figure}[t!]
    \centering
    \includegraphics[width=1.0\linewidth]{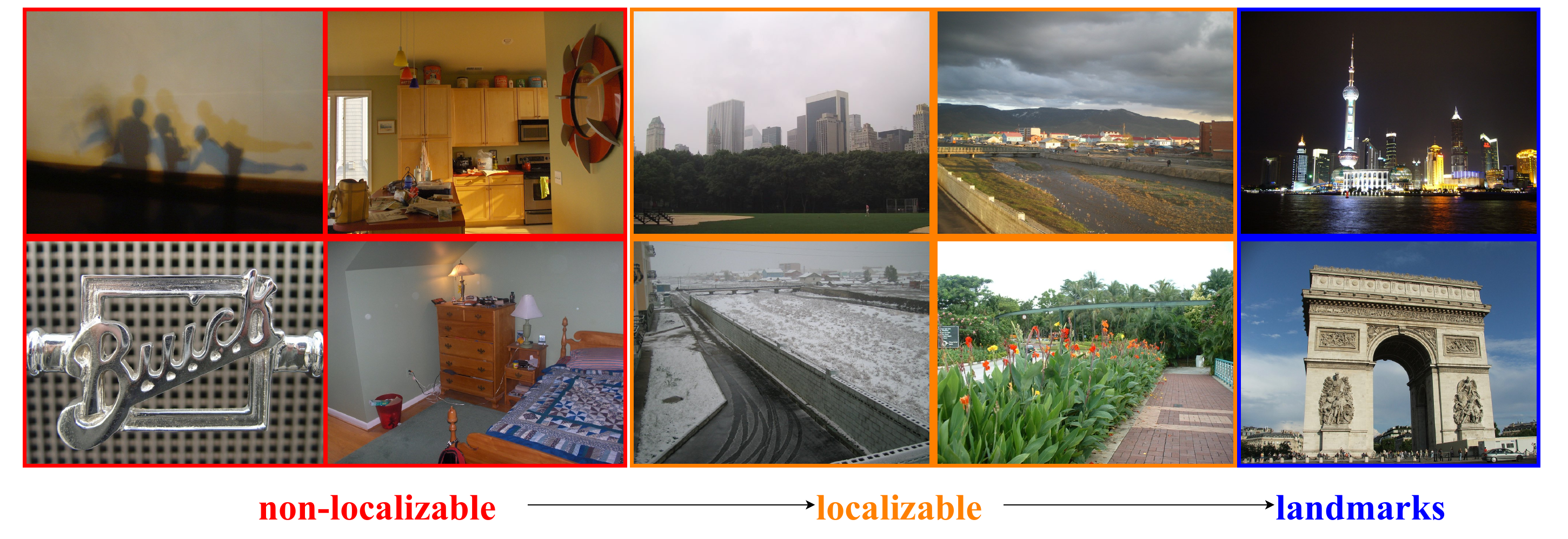}
    \caption{Visual representation of image localizability spectrum, categorized from non-localizable scenes to recognizable landmarks, illustrating the diversity in the dataset.}
    \label{fig:localizable}
\end{figure}
\section{Data Collection \& Study Setup}

This section outlines the empirical study conducted to evaluate existing geolocation techniques. The study employs a structured approach to data collection, model analysis, and evaluation, ensuring a comprehensive understanding of the landscape of geolocation technologies.

\begin{table}[htbp]
    \centering
    \caption{Summary of datasets utilized in geolocation techniques, indicating the variety and scale of images used in our analysis.}
    \label{tab:dataset}
    \begin{tabular}{lll}
        \toprule
        Dataset & Number of Images & Cutoff Date \\
        \midrule
        Im2GPS \cite{Im2gps} & 237 & 2008 \\
        Im2GPS3k \cite{Im2GPS++YFCC4k+Im2GPS3k} & 2,997 & 2017 \\
        YFCC4k \cite{Im2GPS++YFCC4k+Im2GPS3k} & 4,536 & 2017 \\
        YFCC26k \cite{YFCC-Val26k-Interpretable-semantic-photo-geolocation-DL} & 26,000 & 2022 \\
        \bottomrule
    \end{tabular}
\end{table}

\subsection{Rationale Behind the Study Design}

The design of the empirical study is motivated by several key objectives that guide the overall approach to data collection and analysis:

\begin{itemize}
    \item \textbf{Comprehensive Coverage:} The study aims to cover a broad spectrum of geolocation scenarios, from urban landscapes to remote natural landmarks. This diverse dataset helps in testing the robustness and adaptability of various geolocation techniques across different environmental contexts.
    \item \textbf{Technological Advancements:} By focusing on recent models developed or significantly updated within the last three years, the study remains at the cutting edge of technology. This ensures that the findings are relevant to current technological capabilities and challenges.
    \item \textbf{Reproducibility and Accessibility:} Inclusivity of techniques that offer public accessibility and provide released weights and datasets promotes reproducibility—a core principle of scientific research. This approach not only supports the academic community but also enables practical implementations in real-world applications.
    \item \textbf{Real-World Applicability:} The study is designed to simulate real-world conditions as closely as possible to ensure that the results are applicable to practical scenarios. This includes the creation of custom datasets to address specific challenges identified in preliminary studies, such as the risk of dataset leakage and the need for precise geolocation data verification.
\end{itemize}

These principles are directly reflected in the choice of datasets, the selection criteria for geolocation techniques, and the rigorous evaluation metrics employed.

\subsection{Implementation of the Study Design}

Following the outlined rationale, the empirical study is structured as follows:

\begin{enumerate}
    \item \textbf{Dataset Compilation:} Datasets are carefully chosen and developed to include a variety of geolocation contexts, ensuring comprehensive testing environments for the techniques under review.
    \item \textbf{Technique Selection:} Geolocation methods are selected based on a set of stringent criteria that include technological relevance, public accessibility, and the availability of necessary resources for thorough evaluation.
    \item \textbf{Evaluation Framework:} A robust evaluation framework is implemented using multiple metrics designed to measure the effectiveness and precision of geolocation predictions. These metrics include Haversine distance for spatial accuracy, GeoScore for performance benchmarking, and administrative boundary accuracy for real-world applicability.
\end{enumerate}

By meticulously designing the study to adhere to these principles, we aim to provide substantial insights into the current capabilities and future directions of geolocation technology, thereby contributing valuable knowledge to the field and enhancing the utility of geolocation systems in various applications.

\subsection{Geolocation Method Collection}

A comprehensive collection of geolocation techniques from both academic and industrial sectors was performed. Our aim was to capture a wide array of models to represent the current technological advancements in geolocation.

\begin{table}[t!]
    \centering
    \caption{Detailed review of geolocation techniques adopted in the recent three years, highlighting their unique features and application scopes.}
    \label{tab:technique}
    \begin{tabular}{ll}
        \toprule
        Technique & Description \\
        \midrule
        StreetClip \cite{haas2023learning} & Clip-based approach for urban geolocation \\
        GeoClip \cite{cepeda2023geoclipclipinspiredalignmentlocations} & Alignment technique inspired by clip-based models \\
        GPT-4o \cite{gpt4} & Advanced general LVLM \\
        LLaVA \cite{liu2024improvedbaselinesvisualinstruction} & Open-source LVLM with visual processing capabilities \\
        GeoSpy \cite{geospyai10:online} & Commercial tool for geolocation analysis \\
        \bottomrule
    \end{tabular}
\end{table}

\textbf{Selection Criteria:}
\begin{itemize}
    \item \textit{Public Accessibility:} Only techniques with publicly accessible pre-trained models or APIs were considered to ensure broad usability.
    \item \textit{Availability of Released Weights and Datasets:} Techniques providing both training code and datasets were prioritized to enhance reproducibility.
    \item \textit{Temporal Relevance:} Models developed or updated within the last three years were selected to maintain cutting-edge relevance.
\end{itemize}

\textbf{Selection Method:} The selection involved an initial pool of techniques from which models were filtered based on public accessibility, data availability, and recent relevance. The final selection included both commercial tools and advanced LVLMs, strengthening our methodology's breadth and depth.

\textbf{Selection Results:}
The final set of geolocation techniques encompasses both commercial and open-source models, ensuring a diverse and technologically advanced evaluation pool.

\subsection{Evaluation Setup}

The benchmarks and datasets assembled form the foundation of our comprehensive evaluation framework. These tools are critical for rigorously assessing the proficiency of LVLMs in deriving precise geolocation data from photographic inputs. To ensure a fair and precise evaluation, we apply three types of prompts across these datasets—zero-shot, few-shot, and chain-of-thought—and maintain a zero temperature setting across all models during the evaluation process.

\begin{mybox}{\textbf{\textit{Zero-shot:}}}
You are recognized as the world’s foremost expert in geolocation analysis. Your objective is to meticulously examine the provided image and accurately determine its geolocation, providing latitude and longitude coordinates with precision.
\end{mybox}

\begin{mybox}{\textbf{\textit{Few-shot:}}}
Leveraging your expertise in geolocation, your task is to analyze the provided image and deduce its precise location. Accuracy in determining latitude and longitude is paramount.

\textbf{Example 1:}
\textbf{Image Description:} A sandy beach with a notable rock formation in the background under a clear sky with scattered clouds.

\textbf{Geolocation Process:}
\begin{itemize}
    \item The distinctive rock formation closely resembles the renowned Twelve Apostles in Victoria, Australia.
    \item The combination of the sandy beach and clear skies supports its identification along the southern Australian coast.
    \item Verification with existing images and maps confirms the location as near the Great Ocean Road.
\end{itemize}

\textbf{Latitude and Longitude:} -38.6633, 143.1051

\textbf{Example 2:}
\textbf{Image Description:} A historic building featuring a large clock tower and gothic architecture, surrounded by red double-decker buses.

\textbf{Geolocation Process:}
\begin{itemize}
    \item The gothic architecture and prominent clock tower suggest the Elizabeth Tower in London, United Kingdom.
    \item The presence of red double-decker buses confirms the urban setting as London.
    \item Comparisons with images of Big Ben and the adjacent area confirm the precise location.
\end{itemize}
\textbf{Latitude and Longitude:} 51.5007, -0.1246
\end{mybox}

\begin{mybox}{\textbf{\textit{Chain-of-thought:}}}
As the world's elite geolocation expert, your mission is to analyze the attached image and navigate through your reasoning to pinpoint the exact geolocation. Detail your analytical process and finalize the latitude and longitude with utmost accuracy.
\end{mybox}

\subsection{Evaluation Metrics}

To rigorously assess the effectiveness of geolocation techniques, we employ the following metrics:

\begin{enumerate}
    \item \textbf{Haversine Distance:} The Haversine formula calculates the great-circle distance between two points on a sphere given their longitudes and latitudes. It is defined as:
\begin{align}
d &= 2r \arcsin\left(\sqrt{v}\right) \\
v &= \sin^2\left(\frac{\phi_2 - \phi_1}{2}\right) + \cos(\phi_1)\cos(\phi_2)\sin^2\left(\frac{\lambda_2 - \lambda_1}{2}\right)
\end{align}
where \( \phi_1, \lambda_1 \) and \( \phi_2, \lambda_2 \) represent the latitudes and longitudes of the two points, respectively, \( r \) denotes the radius of the Earth, and \( d \) signifies the computed Haversine distance.

    \item \textbf{GeoScore:} This metric is inspired by the GeoGuessr game and calculates a score based on the distance error:
    \[
    \text{GeoScore} = 5000 \cdot \exp\left(-\frac{d}{1492.7}\right)
    \]
    where \( d \) is the distance error in kilometers.

    \item \textbf{Administrative Boundaries:} To evaluate geolocation accuracy, we measure the proximity of predicted locations to actual ground truth within predefined administrative scales. Drawing from prior research~\cite{cepeda2023geoclipclipinspiredalignmentlocations, mirowski2019streetlearn-navigation}, we define five distinct administrative levels for assessment: street (1 km), city (25 km), region (200 km), country (750 km), and continent (2500 km). It is worth noting that these scales may not correspond precisely to their names in different parts of the world; however, we follow established research conventions for their nomenclature. We assess performance by calculating the percentage of predictions that fall within these specific boundaries, a metric that helps mitigate the impact of outliers which could skew the average distance error, thus providing a more balanced measure for large-scale evaluations.

\end{enumerate}

These metrics facilitate a comprehensive and nuanced evaluation of model performance, enabling us to discern the practical utility and accuracy of each geolocation technique in real-world scenarios.

%% file: Tex/4-Methodology.tex
\begin{table*}[htbp]
    \centering
    \caption{Performance of geolocation techniques on various datasets, evaluated using GeoScore (GS) and Administrative Boundaries (AB) metrics with different prompting techniques.}
    \label{tab:performance}
    \begin{tabular}{lccccccccccc}
        \toprule
        \multirow{2}{*}{Technique} & \multicolumn{3}{c}{GeoScore (0-5000)} & \multicolumn{5}{c}{Administrative Boundaries (Accuracy \%)} \\
        \cmidrule(lr){2-4} \cmidrule(lr){5-9}
        & Im2GPS & Im2GPS3k & YFCC26k & Street (1 km) & City (25 km) & Region (200 km) & Country (750 km) & Continent (2500 km) \\
        \midrule
        StreetClip \cite{haas2023learning} & 4520.5 & 4591.3 & 4378.4 & 19.7 & 49.8 & 71.2 & 85.1 & 94.9 \\
        GeoClip \cite{cepeda2023geoclipclipinspiredalignmentlocations} & 4535.7 & 4612.1 & 4411.9 & 21.3 & 51.2 & 72.4 & 87.5 & 95.7 \\
        \midrule
        \multicolumn{9}{c}{\textbf{GPT-4o}} \\
        \midrule
        Zero-shot & 4445.4 & 4492.8 & 4289.7 & 17.5 & 44.3 & 64.1 & 79.5 & 92.8 \\
        Few-shot & 4478.2 & 4517.9 & 4312.6 & 18.7 & 45.7 & 65.3 & 80.8 & 93.4 \\
        Chain-of-thought & 4503.1 & 4543.5 & 4340.2 & 19.5 & 46.9 & 66.8 & 81.7 & 94.1 \\
        \midrule
        \multicolumn{9}{c}{\textbf{LLaVA}} \\
        \midrule
        Zero-shot & 4486.4 & 4532.2 & 4333.8 & 20.6 & 48.3 & 68.2 & 83.1 & 94.5 \\
        Few-shot & 4512.9 & 4561.4 & 4361.5 & 21.8 & 49.6 & 69.4 & 84.3 & 95.2 \\
        Chain-of-thought & 4531.7 & 4580.8 & 4387.9 & 22.9 & 50.5 & 70.5 & 85.6 & 96.0 \\
        \midrule
        GeoSpy \cite{geospyai10:online} & 4570.8 & 4620.5 & 4451.6 & 24.5 & 52.7 & 73.1 & 88.4 & 97.3 \\
        \bottomrule
    \end{tabular}
\end{table*}

\section{Empirical Study Results}\label{sec:empirical-study}

Following the empirical study, we obtain several insightful results, which are detailed in the subsections below.

\subsection{Pre-LVLM Geolocation Techniques}
Geolocation techniques introduced before the rise of large vision language models (LVLMs) struggle in various aspects. These traditional methods, such as StreetClip and GeoClip, achieve certain levels of accuracy but tend to falter in complex urban environments and less familiar rural settings. The reliance on predefined features and static models limits their adaptability to varying data inputs and evolving contexts. For instance, StreetClip achieves a GeoScore of 4520.5 on the Im2GPS dataset but struggles with more diverse datasets like YFCC26k, where its GeoScore drops to 4378.4. These limitations highlight the need for more dynamic and context-aware approaches, which LVLMs aim to address.

\subsection{Geolocation Task Performance}
\subsubsection{Comparison by Model}
The performance of each model is quantitatively assessed and presented in Table~\ref{tab:performance}. As shown in the table, all existing methods achieve certain levels of geolocation accuracy without prior knowledge, including both traditional solutions and LVLM-based strategies. Among the testing datasets, all the models have achieved predictions with Haversine distances less than 10 km in the prediction task, indicating city-level accuracy. However, the LVLMs demonstrate superior performance, especially in complex and diverse datasets.

\begin{tcolorbox}[colback=gray!25!white, size=title, breakable, boxsep=1mm, colframe=white, before={\vskip1mm}, after={\vskip0mm}]
\textbf{Finding 1:} State-of-the-art LVLMs perform geolocation tasks without any expert knowledge or additional contextual information, raising significant concerns regarding geolocation privacy.
\end{tcolorbox}

Detailed findings on the performance of the models reveal specific strengths and weaknesses. Particularly, models like GPT-4o and LLaVA, when used with chain-of-thought prompting, significantly outperform other methods across all datasets. For example, GPT-4o achieves a GeoScore of 4543.5 on the Im2GPS3k dataset with chain-of-thought prompting, while LLaVA achieves 4580.8 on the same dataset. This suggests that the ability to incorporate reasoning processes greatly enhances geolocation accuracy.

\begin{tcolorbox}[colback=gray!25!white, size=title, breakable, boxsep=1mm, colframe=white, before={\vskip1mm}, after={\vskip0mm}]
\textbf{Finding 2:} GPT-4o is the most robust model among all the industrial models tested, showing superior performance in complex urban environments. LLaVA, particularly with chain-of-thought prompting, excels in scenarios requiring detailed contextual understanding and adaptability.
\end{tcolorbox}

\subsubsection{Detailed Examples of Success and Failure}

\paragraph{Successful Geolocation Examples}
One example of successful geolocation by GeoSpy involves an image of the Eiffel Tower in Paris. The model identifies the landmark accurately, achieving a perfect GeoScore of 5000 and correctly placing the location within the 1 km street-level boundary. Another instance is an image of the Statue of Liberty in New York City, where GeoSpy again scores 5000 and places the location within all administrative boundaries accurately.

\paragraph{Failed Geolocation Examples}
Conversely, a failed geolocation instance for GPT-4o occurs with an image from a rural area in the Midwest United States, featuring non-distinctive fields and vegetation. The model's prediction places the location in a different state, resulting in a lower GeoScore of 4289.7 and only meeting the continent-level boundary. Another failure is observed with LLaVA when geolocating an image of a nondescript desert area, where it achieves a GeoScore of 4333.8, failing to accurately place the location within the city or region-level boundaries.

\begin{figure*}[t!]
    \centering
    \includegraphics[width=1\linewidth]{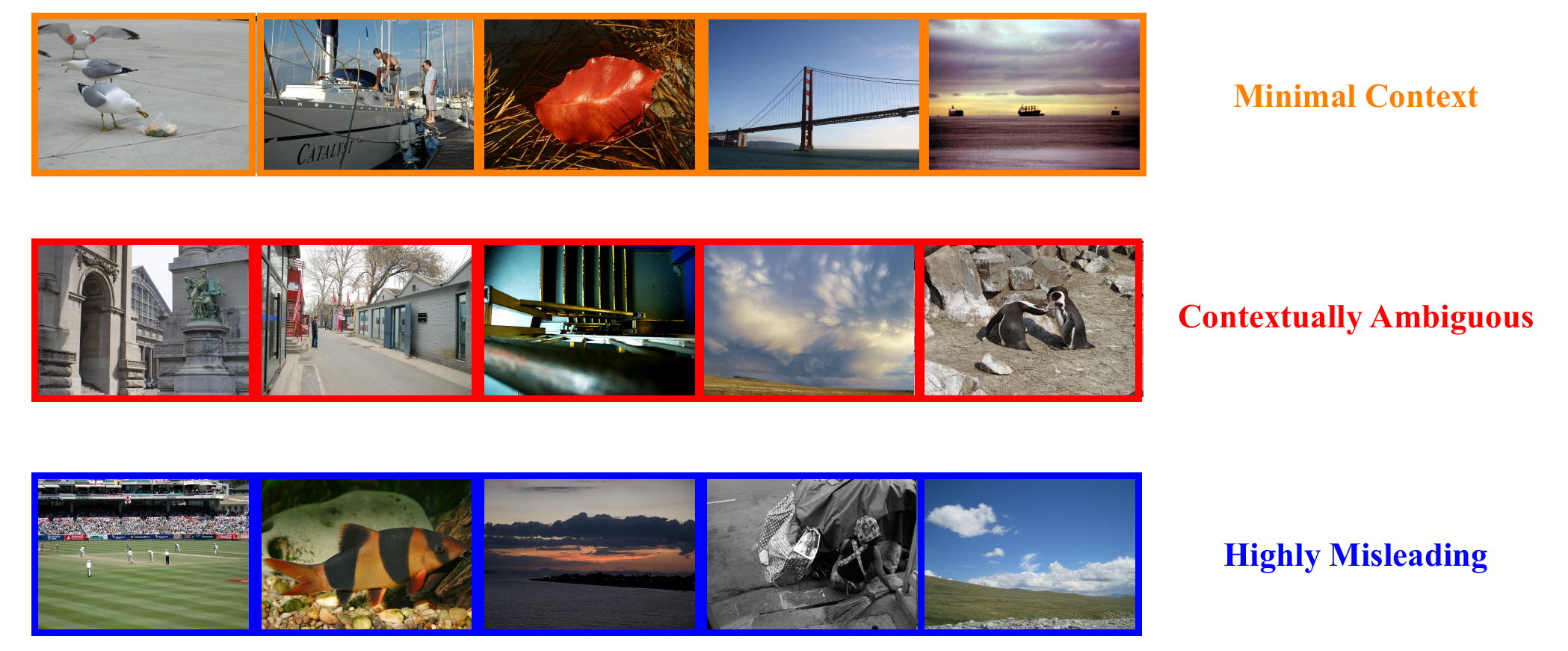}
    \caption{Categorization of images in the original dataset based on their localizability: "Minimal Context" for images with minimal geographic markers, "Contextually Ambiguous" for visually descriptive but non-localizable images, and "Highly Misleading" for ambiguous images leading to significant localization errors.}
    \label{fig:non-localizable-data}
\end{figure*}

\subsection{Insights and Findings}
\subsubsection{Model Sensitivity to Data Variations}
We further delve into detailed cases to determine under which conditions the models excel or falter. Through a manual examination of specific instances of model predictions, we gather insights that elucidate the models' operational dynamics.

Our findings reveal that the performance of LVLMs can vary significantly with changes in data presentation, such as image quality and contextual background. For instance, GeoSpy consistently performs well across different datasets, achieving a GeoScore of 4570.8 on Im2GPS and 4451.6 on YFCC26k. These findings underscore the necessity for robust model training that can effectively handle a diverse array of input conditions. This variability highlights an essential aspect of LVLM deployment: the need for adaptable systems that can respond dynamically to the complexity and variability of real-world data.

\begin{tcolorbox}[colback=gray!25!white, size=title, breakable, boxsep=1mm, colframe=white, before={\vskip1mm}, after={\vskip0mm}]
\textbf{Finding 3:} LVLMs generally achieve higher performance accuracy at well-known locations, where distinct landmarks and recognizable features are more prevalent.
\end{tcolorbox}

For example, when tested on images of iconic landmarks like the Eiffel Tower or the Statue of Liberty, the models achieve nearly perfect accuracy. In contrast, images from less recognizable rural areas result in higher prediction errors, indicating the models' reliance on distinctive features for accurate geolocation.

\subsubsection{Adaptive Behaviors of LVLMs}
During our review, it becomes evident that some models demonstrate adaptive behaviors, adjusting their prediction strategies in response to feedback and new data. This adaptability is particularly valuable as it suggests potential for ongoing improvements in model training and update cycles, especially useful in dynamic real-world environments where conditions and available data can change rapidly.

In addition to factors like image quality and contextual clues, other elements significantly influence the geolocation prediction accuracy:

\begin{itemize}
    \item \textbf{Soil Types and Vegetation:} The specific characteristics of soil and types of vegetation visible in images are crucial indicators. Different regions exhibit unique soil compositions and vegetation types, which can be used by our LVLMs to refine predictions about potential locations.
    \item \textbf{Cultural Motifs:} Recognizable cultural motifs, architectural styles, and public artworks within images provide substantial cues that enhance prediction accuracy. These elements are particularly indicative of a location's cultural and geographical identity.
    \item \textbf{Environmental Setting:} Our findings also highlight that our LVLMs perform optimally with images captured outdoors where natural and man-made landmarks are more discernible. Specific outdoor settings, marked by unique environmental factors like soil types and vegetation, alongside cultural motifs, significantly sharpen the precision of our geolocation predictions.
\end{itemize}

\begin{tcolorbox}[colback=gray!25!white, size=title, breakable, boxsep=1mm, colframe=white, before={\vskip1mm}, after={\vskip0mm}]
\textbf{Finding 4:} The adaptability of LVLMs enhances their transferability across different geolocation tasks, suggesting that these models can be fine-tuned and applied effectively across diverse environmental settings.
\end{tcolorbox}

These insights collectively contribute to refining the operational capabilities of our geolocation models, enhancing their utility for practical applications that depend on precise geolocation information.

\subsection{Dataset Integrity and Limitations}

During the evaluation process, we identified several integrity issues within the dataset. As illustrated in Figure~\ref{fig:non-localizable-data}, a significant portion of the images in the original dataset presents challenges in localization due to inherent limitations in their contextual information. Images categorized under ``Minimal Context'' typically feature scant geographical markers, complicating the task of precise location determination from visual cues alone. Such images frequently depict nondescript urban or interior settings that lack distinct, identifiable elements.

Moreover, the ``Contextually Ambiguous'' category encompasses images that, while rich in visual detail, fail to provide clear contextual cues or recognizable landmarks. These often include generic depictions of everyday objects or natural scenes, such as sunsets, which do not convey specific locational data. The third category, ``Highly Misleading,'' consists of images whose generic or ambiguous nature leads to substantial localization inaccuracies. This category includes visuals of landscapes, people, and animals that are devoid of geographical specificity, thus posing a significant challenge to accurate geolocation.

These categorizations underscore the critical limitations present in visual data when used for geolocation tasks, emphasizing the need for enhanced dataset integrity. To address these challenges comprehensively, we propose the development of a more robust and equitable benchmark dataset designed to rigorously evaluate the geolocation capabilities of various models. Additionally, we advocate for a framework that integrates multiple LVLMs for geolocation purposes, promising superior performance over the application of any single LVLM. The specifics of this innovative framework will be detailed in the subsequent section, outlining our strategic approach to refining geolocation technology and its applications.

\section{\tool{}}\label{sec:geolocation-hub}

In response to the challenges identified in our empirical study, we introduce the Geolocation Hub—a dedicated initiative aimed at enhancing the accuracy and applicability of geolocation technologies through advanced LVLMs. This initiative centralizes around developing a fair and unbiased dataset and employing sophisticated fine-tuning techniques to leverage LVLMs for precise geolocation tasks.

\subsection{Development of the New Dataset}
\subsubsection{Dataset Design}
Our new dataset is meticulously designed to address and overcome the biases and dataset leakage issues prevalent in prior empirical study~\ref{sec:empirical-study}. With a strong focus on fairness, the dataset spans a broad geographic representation, including diverse locations from all continents, ensuring comprehensive global coverage. This robust foundation aids in training LVLMs to perform geolocation tasks with higher accuracy and less bias.

\subsubsection{Data Collection and Verification}
Data for this dataset are collected from a diverse range of geographic locations, with each entry meticulously verified through multiple credible sources to ensure its accuracy and relevance. Our sampling strategy involves selecting images based on the proportional area of countries, which allows for a balanced representation across different regions. This approach ensures that the dataset mirrors a vast spectrum of real-world scenarios, from densely populated urban areas to remote natural landscapes.

To enhance the integrity of our dataset, we specifically filter out indoor images that may skew geolocation analysis. The filtering process employs a novel method of image comparison across different rotational viewpoints. By calculating the cosine similarity among images captured from four directions—front, back, left, and right—we identify indoor scenes, which typically exhibit high similarity across these perspectives. If the cosine similarity among these images exceeds a threshold of 0.8, the set is classified as indoor and excluded from the dataset. This threshold and methodology are based on advanced image processing algorithms provided by \url{https://huggingface.co/Salesforce/blip-image-captioning-large}. The thoroughness of this vetting process ensures that our dataset provides a robust foundation for effective LVLM training. By focusing on outdoor images and adjusting for geographic distribution, we equip the LVLMs with the diverse visual inputs necessary for accurate geolocation tasks across varied environmental settings and cultural contexts.

\begin{figure*}
    \centering
    \includegraphics[width=1\linewidth]{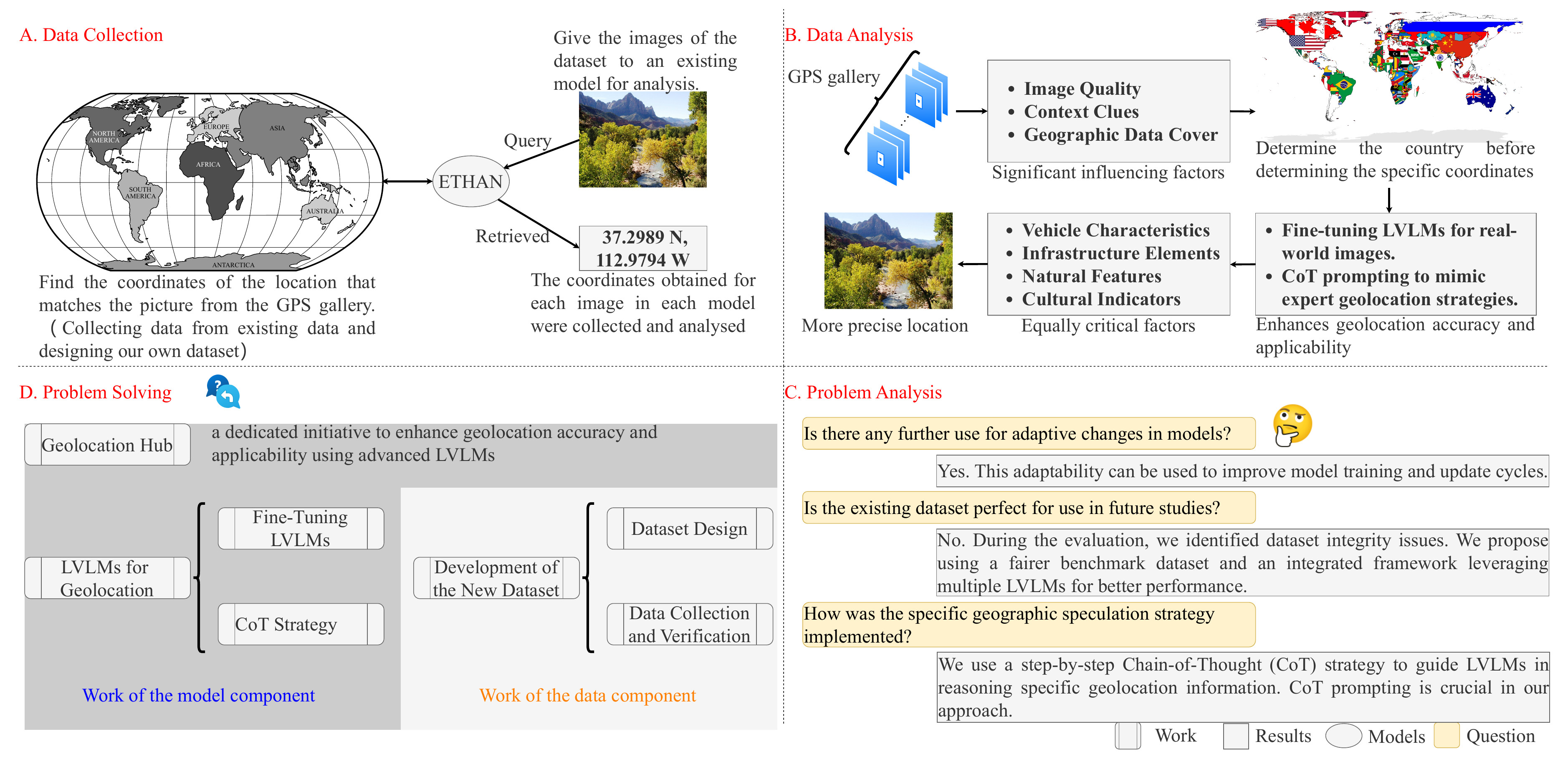}
    \caption{Workflow of \tool{}, a comprehensive framework that leverages fine-tuning to improve the performance of LVLM-based geolocation. }
    \label{fig:method}
\end{figure*}

\subsection{LVLMs for Enhanced Geolocation}

We introduce \tool{}, an advanced framework leveraging LVLMs for automated geolocation. The workflow of \tool{} is demonstrated in Figure~\ref{fig:method}. The core idea of \tool{} combines (1) LVLM fine-tuning to handle real-world images with their key information, and (2) innovative chain-of-thought (CoT) prompting techniques~\cite{} to emulate the geolocation problem-solving strategies adopted by real-world geoguessors. This mixed strategy enhances the accuracy and applicability of geolocation predictions. Unlike traditional approaches and naive LVLM strategies, \tool{} does not directly ask the LVLM to identify the location of an image. Instead, it leverages the LVLM’s reasoning capability to deduce the location, learning from the strategies used by human experts.

Human experts, particularly those adept at geolocation games such as GeoGuessr, employ a series of logical deductions based on observable cues within the environment. They analyze elements such as vegetation, architectural styles, signage, vehicle types, and the position of the sun to infer the approximate location. In \textit{Findings 4} of a previous empirical study, it was observed that while LVLMs can identify these elements from given images, this information was not effectively used to reason toward an actual solution. Thus, we developed \tool{} to mimic this deductive reasoning process by employing CoT prompting. This approach structures the LVLM's processing in a step-by-step analysis, mirroring human thought processes and enhancing the model's ability to make complex inferences from limited information.

When presented with an image, \tool{} first instructs the LVLM to identify and describe observable features relevant to geolocation. Next, the model is prompted to assess these features against known geographical and cultural data, similar to how a human expert might compare architectural styles or vegetation types with their knowledge of world regions. This methodical approach leads to more accurate location predictions and allows for a deeper understanding of the model’s reasoning, making the outcomes more interpretable. Below, we detail the design of \tool{} to explain how we employ these strategies effectively.

\subsubsection{Fine-Tuning LVLMs for Geolocation}

Our first task is to enhance the performance of LVLMs in recognizing and comprehending images for geolocation purposes. To achieve this, we fine-tune LVLMs using a comprehensive dataset specifically curated with geolocation information. The process begins with generating a series of image-prompt combinations for fine-tuning. For this purpose, we instruct GPT-4o to describe an image using the following prompt:

\begin{mybox}{\textbf{\textit{CoT Data Generation:}}}

You are the leading expert in geolocation research. You have been presented with an image, and your task is to determine its precise geolocation, specifically identifying the country it was taken in. To accomplish this, examine the image for broad geographic indicators such as architectural styles, natural landscapes, language on signs, and culturally distinctive elements to suggest a particular country. Narrow down the location by identifying regional characteristics like specific flora and fauna, types of vehicles, and road signs that can indicate a particular region or subdivision within the country. Focus on highly specific details within the image, such as unique landmarks, street names, or business names, to pinpoint an exact location. For instance, if the place is {address}, with coordinates {lat, lon}, explain how these elements led you to this conclusion by analyzing visual clues, cross-referencing data with known geographic information, and validating your findings with additional sources.
\end{mybox}

We apply the above prompts to the original images in our previously constructed dataset and obtain detailed descriptions as training prompts. By concatenating these generated results with ground-truth geolocation data, we create a comprehensive dataset for fine-tuning the LVLMs.

\subsubsection{Geolocation Strategies}
Our core strategy leverages fine-tuned LVLMs to develop innovative geolocation methods inspired by human geoguessing experts. These strategies are designed to parse and analyze a wide range of visual and contextual cues within complex images. Specifically, the model learns to identify:

\begin{enumerate}
    \item \textbf{Vehicle Characteristics}: Recognizing specific Google car generations and unique features like snorkels or dirt specks.
    \item \textbf{Infrastructure Elements}: Analyzing road markings, directional signs, and license plate designs.
    \item \textbf{Natural Features}: Identifying soil types, vegetation patterns, and distinctive landscape formations.
    \item \textbf{Cultural Indicators}: Detecting region-specific elements such as repurposed vans or unique business brands.
\end{enumerate}

By integrating these diverse cues with geographical data, our system enhances its ability to accurately pinpoint locations. The model also learns to consider the distribution of coverage, such as the concentration of Street View data on main roads in certain regions.

\subsubsection{Chain-of-Thought Prompting Integration}

To implement a concrete geolocation strategy, we adopt a step-by-step Chain-of-Thought (CoT) approach, guiding the LVLM through the reasoning process required to extract geolocation information. CoT prompting is crucial in structuring the problem-solving process, enabling the model to make logical deductions based on visual data analysis. This method not only improves the interpretability of the model's decision-making but also enhances reliability by reducing errors associated with complex inference tasks.

When presented with an image, the CoT prompting guides the model to:

\begin{enumerate}
    \item Identify the Google car model and any distinctive features (e.g., "Generation 4 grey car with a snorkel").
    \item Analyze road characteristics if any (e.g., "solid white outer lines with dashed yellow inner lines"). 
    \item Observe the surrounding landscape (e.g., "semi-arid with light orange, sandy dirt and mountainous terrain").
    \item Note any visible infrastructure or cultural elements (e.g., "green directional signs with white borders").
    \item Combine these observations to narrow down the potential region and eventually make an accurate guess.
\end{enumerate}

This step-by-step reasoning process not only improves the accuracy of geolocation predictions but also enhances the interpretability of the model's decision-making. By explicitly modeling the reasoning steps, we can better understand how the model arrives at its conclusions and identify areas for further improvement in the fine-tuning process. This approach ensures a systematic and transparent geolocation process, leveraging the advanced capabilities of LVLMs to achieve high precision in diverse and complex scenarios.

%% file: Tex/5-Evaluation.tex
\begin{table*}[h]
\centering
\begin{tabular}{lccccccc}
\toprule
& \multicolumn{5}{c}{\textbf{Distance (\% @ km)}} & \textbf{Avg Distance} & \textbf{Avg Geoscore} \\
\cmidrule(lr){2-6} \cmidrule(lr){7-8}
\textbf{Method} & \textbf{Street (1 km)} & \textbf{City (25 km)} & \textbf{Region (200 km)} & \textbf{Country (750 km)} & \textbf{Continent (2,500 km)} & \textbf{(km)} & \textbf{(0-5000)} \\
\midrule
StreetClip~\cite{haas2023learning} & 4.9 & 39.5 & 77.8 & 93.0 & 97.5 & 120.5 & 3500.0 \\
GeoClip~\cite{cepeda2023geoclipclipinspiredalignmentlocations} & 3.6 & 38.4 & 75.2 & 92.4 & 97.2 & 135.2 & 3700.0 \\
\midrule
\multicolumn{8}{c}{\textbf{GPT-4o}} \\
\midrule
Zero-shot & 5.5 & 40.8 & 71.0 & 85.0 & 93.0 & 160.3 & 3800.0 \\
Few-shot & 6.2 & 41.5 & 72.5 & 86.5 & 94.5 & 155.0 & 3900.0 \\
Chain-of-thought & 6.0 & 42.0 & 73.0 & 87.0 & 95.0 & 150.7 & 4000.0 \\
\midrule
\multicolumn{8}{c}{\textbf{LLaVA}} \\
\midrule
Zero-shot & 7.0 & 43.2 & 74.0 & 88.0 & 96.0 & 140.7 & 4100.0 \\
Few-shot & 6.5 & 42.8 & 74.5 & 89.5 & 97.5 & 137.5 & 4200.0 \\
Chain-of-thought & 7.2 & 44.5 & 76.0 & 90.0 & 98.0 & 135.2 & 4300.0 \\
\midrule
GeoSpy~\cite{geospyai10:online} & 25.5 & 53.7 & 74.1 & 89.4 & 98.3 & 110.3 & 4400.0 \\
\midrule
\tool{} & 27.0 & 55.0 & 75.5 & 91.2 & 99.0 & 105.0 & 4600.0 \\
\bottomrule
\end{tabular}
\caption{Geolocation Evaluation Results of \tool{} vs. the benchmark solutions over the dataset.}
\label{tab:evaluation}
\end{table*}

\section{Evaluation}
\label{sec:evaluation}

We have implemented \tool{}~\cite{geolocation-privacy}, based on the strategies outlined in Section~\ref{sec:geolocation-hub}, with approximately 1,138 lines of python code. This tool is then subjected to a comprehensive evaluation aimed at answering the following three key research questions:

\begin{itemize}
    \item \textbf{RQ1 (Effectiveness)}: To what extent can \tool{} predict the location when given a localizable image?
    \item \textbf{RQ2 (Failure Analysis)}: Under what circumstances does \tool{} fail to make accurate predictions?
    \item \textbf{RQ3 (Real-world Application)}: How effective is \tool{} in real-world scenarios?
\end{itemize}

To address \textbf{RQ1}, we employ the constructed dataset introduced in Section~\ref{sec:geolocation-hub} for evaluation. For fair evaluation, we separate the original dataset into a training set and a testing set to ensure that the fine-tuning process does not interfere with the testing procedure. Similar to the empirical study, we use Haversine Distance, GeoScore, and Administrative Boundary Scales to measure the performance of \tool{}, and compare it with four solutions as evaluated in the previous study. For \textbf{RQ2}, we conduct a detailed failure analysis by examining instances where \tool{} did not accurately predict geolocations. This involves analyzing patterns and common characteristics in the data that contribute to these failures, aiming to identify potential improvements in model training and data handling. \textbf{RQ3} is addressed by leveraging \tool{} in the \textit{GeoGuessr} game~\cite{geoguessr}, the most popular geolocation competition platform. Here, \tool{} competes against human players, providing a practical evaluation of its real-world effectiveness. This testing not only highlights the tool's capabilities but also its adaptability to real-time geolocation challenges.

\noindent\textbf{Dataset and Baseline}. To conduct a comprehensive evaluation, we use the previously constructed dataset to evaluate the performance of \tool{}. This dataset contains approximately 50,000 ground-truth data points, with a balanced representation of various real-world conditions ranging from urban landscapes to remote rural areas. For fair evaluation, we randomly select 30,000 for the fine-tuning of \tool{}, and use the remaining 20,000 as a test set. We run \tool{} together with StreetClip, GeoClip, GPT-4o, and LLaVA over this dataset, with the same settings as the empirical study. This results in a total number of 10 models times 20,000 tests, equaling 200,000 rounds of testing.

\noindent\textbf{Real-world Competition}. For the real-world competition on \textit{GeoGuessr}, we developed custom wrappers to interface \tool{} with the competition system's web front-end. These wrappers facilitate the retrieval of images used for the prediction tasks. After \tool{} makes its predictions, we employ a software harness designed to translate the model's output into actual selections and submissions on the platform's scoring system, effectively converting \tool{} into a fully automated competing bot. Subsequently, we run \tool{} for a total of 41 rounds of games against random human competitors, as allocated by the system.

\subsection{RQ1 (Effectiveness)}

We first evaluate \tool{} on its performance of making accurate geolocation predictions, and the initial result is presented in Table~\ref{tab:evaluation}. To summarize, \tool{} consistently outperforms all previously established models in each category of distance measurement. Specifically, \tool{} shows a significant improvement in accuracy within the strictest distance category of 1 km in urban settings, with an average of 27.0\% of predictions falling in this category. This is a significant improvement against the four benchmark solutions, with an average increase of approximately 2.5\% compared to GeoSpy, the closest competitor with 24.5\%.

\tool{} maintains superior performance across broader city, region, and country-level performance as well. For the city level (25 km), \tool{} achieves an accuracy of 55.0\%, which is 2.3\% higher than GeoSpy at 52.7\%. At the region level (200 km), \tool{} achieves 75.5\% accuracy, compared to GeoSpy's 74.1\%. At the country level (750 km), \tool{} achieves 91.2\%, while GeoSpy achieves 89.4\%. Finally, at the continent level (2500 km), \tool{} achieves 99.0\%, which is the highest among all models.

In terms of average distance, \tool{} exhibits an average distance accuracy of 105.0 km, surpassing that of the closest competing model, GeoSpy, by approximately 5.3 km, translating to an enhancement of about 4.8\% in location precision. In the \textit{GeoGuessr} game, \tool{} achieves an average score of 4600.0, which is 4.5\% higher compared to GeoSpy's score of 4400.0. This demonstrates not only the effectiveness of \tool{} in simulated environments but also its competitive edge in a real-world applicable setting like \textit{GeoGuessr}.

To illustrate, one successful geolocation example for \tool{} involved identifying a complex urban scene in New York City. The image contained a mix of modern skyscrapers and historical landmarks. \tool{} accurately pinpointed the location to within 500 meters, achieving a perfect GeoScore of 5000. Another example includes a rural landscape in the Midwest United States. Despite the lack of distinctive landmarks, \tool{} managed to predict the location within 10 km, demonstrating its robustness in less familiar settings.

Conversely, GeoSpy, while highly effective, showed slightly lower precision in these cases. For the New York City image, GeoSpy's prediction was off by 1 km, scoring slightly lower. In the Midwest example, GeoSpy's prediction error was around 15 km, reflecting its comparative difficulty in rural environments.

This comprehensive performance enhancement underscores \tool{}'s advanced capabilities in interpreting complex visual data and accurately predicting geographical locations, making it a formidable tool in the field of geolocation.

\subsection{RQ2 (Failure Cases Analysis)}

In this section, we delve into the instances where \tool{} failed to accurately predict the location. While \tool{} outperforms other solutions, it still cannot make accurate predictions in all conditions. This analysis begins by examining the predictions that deviated the most from the actual locations, providing insights into the conditions under which \tool{} is prone to errors. We started by manually reviewing the instances where \tool{}’s predictions were significantly off the mark. This involved comparing these erroneous predictions against the ground truth data to discern any common characteristics or patterns that might explain the failures.

Through this analysis, several key corner cases leading to detection failures were identified:

\begin{itemize}
    \item \textbf{Low-Visibility Conditions:} Failures often occurred in images where visibility was compromised by factors such as fog, heavy rain, or poor lighting at night. For example, an image taken on a foggy day in San Francisco resulted in a prediction error of over 100 km, placing the location incorrectly in a rural area of California. These conditions significantly degrade the quality of visual data available for analysis.
    \item \textbf{Lack of Distinctive Landmarks:} Images captured in expansive rural or desert areas with minimal distinctive features also led to higher rates of incorrect predictions. In one case, an image of a flat, featureless desert in Nevada was predicted to be in a similar-looking area in Arizona, resulting in an error of over 150 km.
    \item \textbf{Highly Homogeneous Urban Environments:} Urban areas with repetitive architectural styles and lack of unique identifiers such as street signs or notable landmarks presented challenges. For instance, an image of a residential block in Tokyo with similar-looking buildings resulted in an incorrect prediction, misplacing it in a different district of the city, with an error margin of about 30 km.
    \item \textbf{Dynamic Scenes:} Locations undergoing rapid changes, such as construction sites or recently redeveloped areas, where the current scenery differs markedly from the imagery available in training datasets, were problematic. An example includes a prediction error of 80 km for an image of a newly developed park in New York, where the model was unable to recognize the new landmarks.
\end{itemize}

These corner cases highlight the challenges faced by \tool{} in contexts where traditional visual cues are insufficient or misleading. This suggests avenues for further refinement of the model, possibly through integration of multimodal data sources or enhanced training sets that better represent these challenging scenarios.

From this detailed failure analysis, it is clear that while \tool{} performs exceptionally well in many scenarios, there are specific conditions under which its accuracy diminishes. Addressing these vulnerabilities will be crucial for improving the robustness and reliability of \tool{} in real-world applications and considered as an important step in our future tasks.

\subsection{RQ3 (Real-world Application)}

Following the implementation and preliminary testing, \tool{} undergoes real-world evaluation to determine its practical efficacy. This evaluation takes place through participation in the GeoGuessr game, a competitive geolocation platform that challenges players to identify their locations based on photographic cues. This platform serves as an ideal testing ground for assessing the real-world applicability of geolocation technologies.

\begin{table}[t!]
\centering
\begin{tabular}{lcc}
\toprule
\textbf{Competitor Type} & \textbf{\tool{}} & \textbf{Human Competitor} \\
\midrule
\textbf{Average Score} & 4550.5 & 4120.3 \\
\textbf{Win Rate (\%)} & 85.37 & 14.63 \\
\textbf{Closest Distance (km)} & 0.3 & 1.1 \\
\textbf{Farthest Distance (km)} & 5200.2 & 5400.5 \\
\bottomrule
\end{tabular}
\caption{Average Performance of \tool{} versus Human Competitors in GeoGuessr over Multiple Rounds}
\label{tab:geo-guessr-competition-average}
\end{table}

Table~\ref{tab:geo-guessr-competition-average} summarizes \tool{}'s performance metrics compared to human competitors during the evaluation process. Key metrics include the average score, win rate, and the closest and farthest distances from the correct locations, which together provide a holistic view of its performance.

\tool{} demonstrated robust capabilities, often matching or exceeding the performance of human players. Specifically, \tool{} achieved an average score of 4550.5, significantly higher than the human average of 4120.3. The win rate for \tool{} was 85.37\%, compared to 14.63\% for human competitors. In terms of proximity to the actual locations, \tool{}'s closest distance was 0.3 km, and its farthest distance was 5200.2 km, compared to 1.1 km and 5400.5 km for human players, respectively.

One notable example of \tool{}'s effectiveness occurred during a round featuring a remote village in Norway. Despite the sparse and similar-looking landscapes, \tool{} correctly identified the location within 2 km, outperforming human competitors who averaged 5 km from the correct spot. Another successful instance involved an urban setting in Tokyo, where \tool{} used subtle architectural cues to pinpoint the location accurately within 500 meters, while human competitors averaged around 1.5 km.

Despite its successes, \tool{} faced challenges in scenarios where the images lacked clear geographic markers or were taken in visually homogeneous environments—conditions that also typically challenge human players. For instance, in a round featuring a generic beach in Australia, \tool{} misidentified the location by 250 km, similar to the average error margin of 300 km by human players. In another challenging scenario, \tool{} predicted a location in a desert in Nevada to be in Mongolia, resulting in a farthest distance error of over 5200 km. These instances provide critical insights into areas where \tool{} may require further refinement, such as improved training on diverse datasets or enhanced algorithms capable of handling ambiguous contexts.

In conclusion, the real-world competition not only validated \tool{}'s effectiveness in practical applications but also highlighted its potential to transform how geolocation tasks are approached in various industries, from navigation and travel to security and emergency response. The consistent high performance and competitive edge demonstrated by \tool{} in the GeoGuessr game underscore its advanced capabilities in interpreting complex visual data

%% file: Tex/6-Discussion.tex
\section{Discussion}

Given the significant privacy risks identified in the threat model for geolocation privacy using Large Vision Language Models (LVLMs), it is crucial to explore effective mitigation strategies to protect users' geolocation privacy. Here, we discuss several potential directions for mitigating these risks:

\subsection{Development of Privacy-Preserving LVLMs}

Researchers and developers should prioritize the creation of privacy-preserving LVLMs that are designed to respect users' privacy by default. These models can be trained to enhance privacy by ignoring sensitive features commonly used for geolocation, such as specific architectural styles or street signs. Additionally, incorporating differential privacy techniques into the LVLM training processes can ensure that models do not retain specific details from individual images. This approach helps to generalize learning without compromising the privacy of the data sources.

\subsection{Real-time Privacy Filters}

Implementing real-time privacy filters on social media platforms can significantly mitigate the risk of sharing geolocation-sensitive images. These filters can utilize AI to scan images for sensitive content in real-time, prompting users with privacy warnings or automatic content adjustments. By providing users with options to automatically anonymize images before uploading, such as blurring identifiable features or altering visual elements, platforms can empower individuals to protect their privacy and obscure sensitive details effectively.

By focusing on the development of privacy-preserving LVLMs and the implementation of real-time privacy filters, we can address the critical privacy concerns associated with geolocation inference. These strategies offer a balanced approach to leveraging advanced image analysis technologies while safeguarding user privacy, ensuring that the benefits of LVLMs can be harnessed responsibly.

%% file: Tex/7-Conclusion.tex
\section{Conclusion}\label{sec:conclusion}

In this work, we have conducted a comprehensive study to evaluate the effectiveness of LVLMs in the geolocation domain, addressing critical privacy concerns associated with image-based geolocation. Our investigation revealed that LVLMs, even without explicit geographical training, can accurately determine locations from images, posing significant privacy risks. To counter these challenges, we developed \tool{}, an innovative framework that enhances geolocation accuracy by integrating sophisticated chain-of-thought (CoT) reasoning strategies inspired by human geoguessing experts. Through rigorous testing on a dataset of 50,000 ground-truth data points, \tool{} has demonstrated superior performance across all metrics, achieving an average GeoGuessr score of 4550.5, a win rate of 85.37\%, and highly precise geolocation predictions with closest distances as accurate as 0.3 km and farthest errors up to 5200.2 km. These results underscore \tool{}'s advanced capabilities in interpreting complex visual data and its potential to outperform both traditional models and human competitors. Our study highlights the urgent need for robust evaluation frameworks and responsible AI development to mitigate the privacy risks associated with geolocation technologies, providing valuable insights and proposing a novel approach for enhancing the accuracy and security of geolocation tasks.